\documentclass[12pt]{article}
\usepackage{graphicx,amsmath}
\usepackage{units}

\newlength{\dinwidth}
\newlength{\dinmargin}
\newcommand{\diff}[1]{\frac{\mathrm{d}#1}{#1}}

\def\lapproxeq{\lower .7ex\hbox{$\;\stackrel{\textstyle                                                    
<}{\sim}\;$}}                                                    
\def\gapproxeq{\lower .7ex\hbox{$\;\stackrel{\textstyle                                                    
>}{\sim}\;$}}                                                    
\def\be{\begin{equation}}                                                    
\def\ee{\end{equation}}                                                    
\def\bea{\begin{eqnarray}}                                                    
\def\eea{\end{eqnarray}}

\def\qq{q\bar{q}}

\def\GeV{\rm GeV}

\def\sh{\hat s}
\def\sh2{{\hat s}^2}
\def\PDF{{\rm PDF}}
\def\CLO{C^{\rm LO}}
\def\CNLO{C^{\rm NLO}}
\def\CNNLO{C^{\rm NNLO}}
\newcommand{\dd}{\mathrm{d}\, }

\begin{document}
%\titlepage

\begin{flushright}                                                    
IPPP/12/93  \\
DCPT/12/186 \\                                                    
\today \\                                                    
\end{flushright} 

\vspace*{0.5cm}

\begin{center}
{\Large \bf Improving the Drell-Yan probe of small $x$ partons}\\ 
\vspace*{0.3cm}
{\Large \bf at the LHC via a $k_t$ cut}
\vspace*{1cm}
                                                   
E.G. de Oliveira$^{a,b}$, A.D. Martin$^a$ and M.G. Ryskin$^{a,c}$  \\                                                    
                                                   
\vspace*{0.5cm}                                                    
$^a$ Institute for Particle Physics Phenomenology, University of Durham, Durham, DH1 3LE \\                                                   
$^b$ Instituto de F\'{\i}sica, Universidade de S\~{a}o Paulo, C.P.
66318,05315-970 S\~{a}o Paulo, Brazil \\
$^c$ Petersburg Nuclear Physics Institute, NRC Kurchatov Institute, Gatchina, St.~Petersburg, 188300, Russia \\          
                                                    
\vspace*{1cm}                                                    
                                                    
\begin{abstract}

We show that the observation of the Drell-Yan production of low-mass lepton-pairs ($M\lapproxeq 20$ GeV) at high rapidities ($Y\gapproxeq 3$) at the LHC can make a {\it direct} measurement of parton distribution functions (PDFs) in the low $x$ region, $x \lapproxeq 10^{-4}$.  We describe a procedure that greatly reduces the sensitivity of the predictions to the choice of the factorization scale and, in particular, show how, by imposing a cutoff on the transverse momentum of the lepton-pair, the data are able to probe PDFs in the important low scale, low $x$ domain.
We include the effects of the Sudakov suppression factor.

\end{abstract}                                                        
\vspace*{0.5cm}                                                    
                                                    
\end{center}

\section{Introduction}

\begin{figure} 
\begin{center}
\includegraphics[height=10cm]{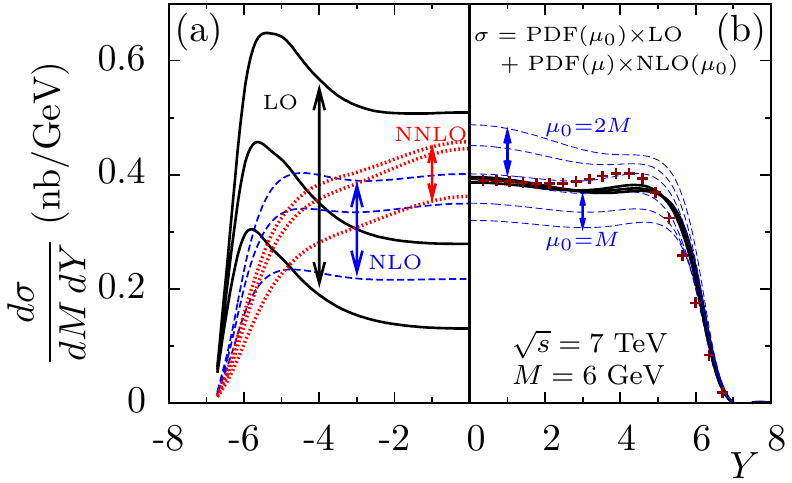}
\vspace{-0.2cm}
\caption{\sf (a) Sensitivity of $M=6$ GeV Drell-Yan $\mu^+\mu^-$ production at the LHC energy of 7 TeV, as a function of rapidity $Y$, to the choice of factorization scale: $\mu_F=M/2,\ M,\ 2M$, at LO, NLO, NNLO, calculated using the programme Vrap \cite{Vrap}, (b) The {\it bold lines} correspond to the choice $\mu_F=\mu_0=1.4M$ which minimizes $\CNLO_{\rm rem}$, and show the stability with respect to the variations $\mu=M/2,\ M,\ 2M$ in the scale of the PDFs convoluted with $\CNLO_{\rm rem}$ -- the $\mu$ dependence is indicated by the symbolic equation at the top of the diagram. The {\it dashed lines} show that the stability (with respect to the variation of $\mu=(M/2,~M, ~2M)$) disappears for other choices of $\mu_0$. The small {\it crosses} are the NNLO result for $\mu=\mu_0=1.4M$ -- note that, here, to see only the effect of the NNLO corrections we have used NLO MSTW partons \cite{MSTW} for both the NLO and NNLO calculations, whereas in (a) the NNLO curves were calculated with NNLO partons. The figure is taken from  \cite{OMR}.}
\label{fig:muF}
\end{center}
\end{figure}

The very high energy of the LHC allows us to probe the parton distribution functions (PDFs) of the proton at extremely small $x$; a region not accessible at previous accelerators. One such process is the Drell-Yan production of low-mass, lepton-pairs, at high rapidity \cite{OMR}, another is C-parity-even quarkonia production \cite{DRS}. Here we study the former process in detail. 
For example, assume, for the moment, Drell-Yan production based on the LO subprocess $\qq \to \gamma^* \to \mu^+\mu^-$. Then the $x$ values of the $q,\bar{q}$ are
\begin{equation}
x_{1,2}~~=~~M e^{\pm Y}/\sqrt{s}~~=~~4.1\times 10^{-2},~~~1.4 \times 10^{-5}
\end{equation}
for a $\mu^+\mu^-$ pair of mass $M=6$ GeV and rapidity $Y=4$, and collider energy $\sqrt{s}=8$ TeV.
Drell-Yan events with these kinematics are in reach of experiments at the LHC, see, for example, Ref. \cite{LHCb}.  However, the predictions of the cross section depend sensitively on the choice of the factorization and renormalization scales. The dependence on the factorization scale, $\mu_F$, is shown in the left half of Fig. \ref{fig:muF}, see also \cite{OMR}.     

Of course, after the summation of many perturbative orders, the final prediction should be much less senstive to the choice of $\mu_F$ that is used to separate the incoming PDFs from the hard matrix element. However, the strong scale dependence at low $x$ comes from the large probability, enhanced by ln$(1/x)$, to emit a new parton in some interval $\Delta \mu_F$. The mean number of partons emitted in the interval $(\mu_F/2,~2\mu_F)$ can be more than 5. So, in principle, we would need to work at more than N$^5$LO perturbative order to provide stability of the prediction. Instead, now that we understand the origin of the problem, we may resum the double logarithms, $(\alpha_s{\rm ln}(1/x){\rm ln}\mu_F^2)^n$, {\it inside} the incoming PDFs by choosing an appropriate scale $\mu_F=\mu_0$ \cite{OMR}. To determine $\mu_0$ we study the detailed structure of the `last cell' in the Feynman diagram, where the incoming PDFs are matched to the hard matrix element. At a relatively low factorization scale the  LO $q\bar{q} \to \gamma^*$ subprocess is overshadowed by
the NLO subprocess $gq \to q\gamma^*$, due to the dominance of the gluon PDF
at low $x$.  
The idea is to use the known NLO result to find the optimal scale such that as much as possible of the ln$(1/x)$
enhanced terms are collected in the LO part of the amplitude; then the predictions will be much more stable to variations of $\mu_F$ in the remaining NLO part. 
This procedure, which allows us to
 resum the most dangerous double logarithms into the incoming PDFs, is outlined in Section \ref{sec:2}.
 
 At first sight we appear to be following the `fastest apparent convergence' (FAC) procedure \cite{FAC}, but there the scale $\mu_F$ is chosen in such a way as to nullify the whole NLO contribution.  However, in the FAC prescription, we can never be sure that the NNLO and higher-order corrections become simultaneously small for the choice $\mu_F=\mu_0$. Our choice of $\mu_0$ is not based on the whole NLO contribution, but on the specific dominant $gq\to \gamma^* q$ subprocess, which has exactly the same Feynman diagram structure as that in LO evolution and so allows us to {\it resum} the leading double logs. This does not exactly nullify all of the NLO contribution. We still have to account for the remaining NLO, NNLO,... terms. However these remaining contributions have no ln$(1/x)$ enhancement, and the convergency is be much better, as demonstrated in \cite{OMR} and Fig. \ref{fig:muF}(b).   If we were to choose a different value of $\mu_0$ the scale dependence caused by the remaining NLO correction is stronger, as shown by the dashed curves in Fig. \ref{fig:muF}(b).

In order to probe the parton distributions at lower scales  we introduce in Section \ref{sec:3} a cutoff, $k_0$, on the transverse momentum, $k_t$, of lepton pair (and implement the above procedure of resumming the large ln$(1/x)$ contributions), in order to study the behaviour of the cross section of Drell-Yan events with $k_t<k_0$. Here one faces another problem -- large higher-$\alpha_s$-order corrections enhanced by logarithms of the ratio $k_0/M$. These double logarithms, of form $(\alpha_s\ln^2(k_0/M))^n$, can be resummed into the Sudakov $T$-factor. 
Thus our approach accounts for (and resums) both possible types of double logarithms which are responsible for the large higher-$\alpha_s$-order 
corrections. As long as the incoming PDFs are known, the remaining NLO corrections only weakly affect the theoretical predictions. 
The cross sections expected from these kinematics are evaluated in Section \ref{sec:4}. Section \ref{sec:5} contains our conclusions.

\section{Procedure to reduce the scale uncertainity   \label{sec:2}}
Basically, the idea in Ref. \cite{OMR} is to write
\be
\sigma^{\rm NLO}(\mu_F)~=~\PDF(\mu_F)\otimes(\CLO + \alpha_s \CNLO_{\rm rem}) \otimes \PDF(\mu_F), 
\label{eq:stab}
\ee
and to note that double counting has to be avoided, since part of the NLO contribution is already included, to leading log accuracy, in the first-order in $\alpha_s$ term in LO DGLAP. The effect of varying the scale  from $m$ to $\mu_F$ in both PDFs of the LO-generated contribution can be expressed as
\be 
\sigma^{\rm LO}(\mu_F)=\PDF(m)\otimes    
 \left(\CLO ~+~\frac{\alpha_s}{2\pi}{\rm ln}\left(\frac{\mu_F^2}{m^2}\right)(P_{\rm left}\CLO+\CLO P_{\rm right})\right)\otimes \PDF(m),
\label{eq:5}
\ee
where the splitting functions
\begin{equation}
P_{\rm right}=P_{qq}+P_{qg}~,~~~~~~~~~P_{\rm left}=P_{\bar{q}\bar{q}}+P_{\bar{q}g}
\end{equation}
act on the right and left PDFs respectively. We may equally well have incoming $\bar{q}$'s in $P_{\rm right}$ and incoming $q$'s in $P_{\rm left}$.
When this LO-generated term is subtracted, the remaining NLO coefficient function $\CNLO_{\rm rem}(\mu_F)$, becomes dependent on $\mu_F$.
As a result, changing $\mu_F$ redistributes the order $\alpha_s$ correction between the LO part ($\PDF\otimes\CLO\otimes\PDF$) and the NLO part ($\PDF\otimes\alpha_s\CNLO_{\rm rem}\otimes\PDF$).

The trick is to choose an appropriate scale, $\mu_F=\mu_0$, so as to minimize the remaining NLO contribution $\CNLO_{\rm rem}(\mu_F)$.  To be more precise, we choose a value  $\mu_F=\mu_0$ such that as much as possible of the `real' NLO contribution (which has a `ladder-like' form and which is strongly enhanced by the large value of $\ln(1/x)$) is included in the LO part where all the logarithmically enhanced, $\alpha_s\ln(1/x)$, terms are naturally collected by the incoming parton distributions.

After we have fixed the scale $\mu_F=\mu_0$ for the LO contribution, we noted above that the remaining NLO contribution starts to depend on $\mu_0$, since to avoid double counting we must subtract from the NLO Feynman diagram that part which is already generated at LO with scale $\mu_0$. In general, the remaining NLO contribution could be evaluated at another scale $\mu_1$: so that we have a term $\PDF(\mu_1)\otimes\alpha_s\CNLO_{\rm rem}(\mu_0)\otimes\PDF(\mu_1)$. Then the remaining NNLO contribution,
$\PDF(\mu_2)\otimes\alpha_s^2\CNNLO_{\rm rem}(\mu_0,\mu_1)\otimes\PDF(\mu_2)$, will depend on $\mu_0$ and $\mu_1$; and so on. Note that unlike the usual form used for the factorized partonic cross section, here at different $\alpha_s$ orders the PDFs are evaluated
at different scales. However, the amount of the inconsistency is always of
one order higher in the perturbative expansion and can therefore be
neglected (or corrected by calculating the  contribution from this one order higher).  In fact, the NNLO results presented in Fig. \ref{fig:muF}(b) were calculated in the conventional way, with $\mu_2=\mu_1=\mu_0$.   

To implement this for Drell-Yan $\mu^+\mu^-$ production at low $x$, we first note that the majority of incoming quarks/antiquarks are produced via the low-$x$ gluon-to-quark splitting, $P_{qg}$. That is, the most important NLO subprocess is $gq\to q\gamma^* $.
The corresponding cross section reads
\begin{align}
\frac{\dd \sigma (gq\to q\gamma^*)}{\dd M^2 \dd t} &= \frac{1}{9} \frac{\alpha^2 \alpha_s}{M^2 {\hat s}^2} \left[-\frac{\hat s}{t} - \frac{t}{\hat s} - \frac{2M^2 u}{\hat st} \right]
\label{eq:exact0}
\\
 &= \frac{\alpha^2 \alpha_s z}{9 M^4} \frac{1}{t} \left[( (1 - z)^2 + z^2) + z^2 \frac{t^2}{M^4} - 2 z^2 \frac{t}{M^2} \right]
 \label{eq:exact}
\end{align}
where $u=M^2-\hat s -t$ and\footnote{Strictly speaking, $z$ is the ratio of the light-cone momentum fraction carried by the `daughter' quark to that carried by the `parent' gluon, $z=x^+_q/x^+_g$.} $z=M^2/s$. The  kinematical upper limit of $|t|$ is 
\be
|t_1|=\hat s-M^2=(1-z)M^2/z,
\ee
On the other hand, the $\alpha_s$ term in the LO-generated DGLAP contribution is
\begin{equation}
\frac{\dd \sigma (gq\to q\gamma^*)}{\dd M^2 \dd t} = \frac{\alpha^2 \alpha_s z}{9 M^4} \frac{1}{t} \left[ z^2 + (1-z)^2 \right].
\label{eq:LLA}
\end{equation}
We integrate the exact result, (\ref{eq:exact}), from $|t_0|$ up to $|t_1|$, and the approximate expression, (\ref{eq:LLA}), from $|t_0|$ to $\mu_0^2$, and choose $\mu_0$ so that the two integrals equal each other; that is
\begin{equation}
z \left[ z^2 + (1-z)^2 \right] \ln\frac{\mu_0^2}{|t_0|} = z \left[( (1 - z)^2 + z^2) \left( \ln \frac{M^2}{|t_0|} + \ln \frac{1 - z}{z}\right)  +  \frac{1 + 3z}{2}(1 - z)   \right],
\label{eq:bal}
\end{equation} 
where the infrared divergency, as $t_0 \to 0$, cancels. After integration of (\ref{eq:bal}) over the incoming gluon,  quark flux\footnote{We take the leading ln$(1/x)$ form of the flux so as not to distort the structure of the double $\alpha_s{\rm ln}(1/x){\rm ln}M^2$ resummation. We find a more flexible form of the flux, with an additional $z^\delta$ factor, has a negligible effect on the value found for $\mu_0$ for a reasonable $\delta < 0.2$.

}, $\dd z/z$, we find that
the optimum scale is given by
\begin{equation}
\ln \frac{\mu_0}{M} = \frac{3}{8},~~~~~{\rm that~is}~~~~~\mu_0 = 1.45M.
\label{eq:nocut}
\end{equation}
Indeed, we can check that the explicit calculation of the pure logarithmic LO
evolution integrated up to $\mu_0=1.45M$ convoluted with $C^{\rm LO}$ reproduces the NLO contribution of this subprocess.
Since in the low-$x$ region $gq \to q\gamma^*$ is the dominant NLO subprocess we anticipate that the choice of $\mu_F=\mu_0=1.45M$ will minimize the NLO contribution, $\CNLO_{\rm rem}(\mu_F)$. 
The insensitivity of the prediction of the NLO cross section, $d\sigma/(dMdY)$, to the choice of scale is shown in the right half of Fig. \ref{fig:muF}. We see that for $\mu_0=1.4M$ the NLO predictions, using the different scales $\mu_F=M/2,\ M,\ 2M$ in the PDFs of the NLO contribution, are virtually indistinguishable from each other.
 
The predictions use the MSTW2008 NLO set of PDFs \cite{MSTW}. The corresponding PDF uncertainty for $Y\gapproxeq 3$ (that is $x_2\lapproxeq 4 \times 10^{-5})$ is such that the error corridor embraces the predictions of the other recent sets of PDFs, see the plot in Fig. 5 of \cite{OMR}. 

Recall that at lower scales the uncertainties of the present low $x$ PDFs strongly increase.
%At first sight, it therefore seems that such data will reveal little information on PDFs. This is not true. 
Pure DGLAP extrapolations in this small $x$ domain are unreliable due to the absence of absorptive, ln$(1/x)$, etc.,.. modifications.  Rather, LHC data for this process will provide {\it direct} measurements of parton distribution functions in this, so far unexplored, low $x$ domain, albeit at the moderately high scale $\mu_F=\mu_0 = 1.45M$. For example, the observations of the production of Drell-Yan $\mu^+\mu^-$ pairs of mass $M=6$ GeV probe PDFs at a scale $Q^2=\mu_0^2 \simeq 75~ \GeV^2$. Clearly, it is desirable to select Drell-Yan data which can measure small $x$ PDFs at lower $Q^2$ scales. Can this be done?

\section{To probe small $x$ PDFs at low scales  \label{sec:3}}

In this Section we investigate the possibility of probing PDFs at smaller $Q^2$ scales by imposing a kinematical cut on the Drell-Yan events.  Naively we may expect that the production of Drell-Yan $\mu^+\mu^-$ pair with a small transverse momentum of the pair $k_t<k_0$ should be described by the PDF measured at scale $\mu\sim k_0$, since the corresponding distribution includes the partons with transverse momentum $p_t<\mu$.  

More precisely the leading order $k_t$ distribution is described by the formula~\cite{DDT}
\be
\label{eq:ddt}
\frac{d\sigma}{dM^2dYdk^2_t}=\frac{\partial}{\partial k^2_t}\left [\PDF(k_t)~\CLO(M)~T^2_q(k_t,M)~\PDF(k_t)\right ]\, ,
\ee
where factor $T_q$ reflects the probability not to emit additional gluons with $p_t>k_t$ during the incoming quark evolution up to scale $M$. This probability is given by the resummation of the virtual ($\propto \delta(1-z)$) loop contributions in the DGLAP equation \cite{KimbMR,MRW}
\be
\label{eq:tq}
 T_q (k_t^2,\mu^2) \equiv \exp \left (-\int_{k_t^2}^{\mu^2}\!\diff{\kappa_t^2}\,\frac{\alpha_S(\kappa_t^2)}{2\pi}\,\int_0^1\!\diff{z}\;z \,P_{qq}(z )\Theta(1-z-\Delta) \right ),
\ee
where the infrared cutoff is
\begin{equation}
\Delta \equiv \frac{k_t}{\mu+k_t}.
\end{equation}
Note also that selecting the events with a cut on the transverse momentum $k_t$ of lepton pair, that is, where the heavy photon has $k_t<k_0$,  we suppress the contribution of the NLO
 $gq\to q\gamma^*$ and $q\bar q\to g\gamma^*$ subprocesses and 
 therefore  
increase the relative importance of the LO contribution. As a consequence, this will {\it decrease} the value we will find for the `optimal' factorization scale, $\mu_0$.  

As a first step we neglect the role of Sudakov $T$-factors and calculate the value of $\mu_0$ as a function of the cutoff $k_t>k_0$,  assuming that $T=1$. This can be done analytically. Then, in the following subsection, we will include the Sudakov effect.

\subsection{Imposing $k_t(\mu^+\mu^-)<k_0$ reduces the optimal scale $\mu_0$  \label{subsec:3.1}}
To explore this possibility, we note that, in
the framework of the collinear approximation, the transverse momentum $k_t$ is generated by  NLO subprocesses, like $q\bar q\to g\gamma^*$ or,  most important for the low $x$ domain, the subprocess $qg\to q\gamma^*$.
Repeating the procedure described in Section \ref{sec:2}, we first calculate the part  
 of the cross section of this last subprocess which satisfies the cut
 $k_t<k_0$. Then we compare it with the cross section generated by the LO
 matrix element convoluted with the last step of the LO DGLAP 
 evolution up to the factorization scale $\mu_F$. We choose 
the value $\mu_F=\mu_0(k_0)$ that, in terms of the LO approach, reproduces the NLO result with the $k_t<k_0$ cut imposed. The result of taking    
the factorization scale $\mu_F=\mu_0(k_0)$ is that, as before, we will again absorb into the LO term the major  part of the NLO contribution; that is, the part that is enhanced by the large values of $\ln(1/x)$. Clearly this `optimal' scale $\mu_0$ now depends on the transverse momentum cutoff $k_0$, and will decrease if we impose a tighter cut.

To determine the dependence of $\mu_0$ on the cutoff $k_0$, we must first rewrite the subprocess cross section (\ref{eq:exact0}) in terms of $k_t^2$, rather than $t$.  The relevant relations are
\begin{equation}
-t = \frac{1}{2} \left[ \hat s - M^2 \pm \sqrt{(\hat s - M^2)^2 - 4 \hat s k_t^2}  \right],~~~~{\rm giving}~~~~
\dd t = \pm \frac{\hat s}{\sqrt{(\hat s - M^2)^2 - 4 \hat s k_t^2}} \dd k_t^2.
\end{equation}
Allowing for both the negative and positive roots, (\ref{eq:exact0}) becomes
\begin{align}
\frac{\dd \sigma (gq\to q\gamma^*)}{\dd M^2 \dd k_t^2} & = \frac{1}{9} \frac{\alpha^2 \alpha_s z}{M^6}  
\frac{z}{\sqrt{(1 - z)^2 - 4 z k_t^2/M^2}} 
\left[ (1 + 2z^2 - 2z) \frac{1-z}{z} \frac{M^2}{k_t^2} + 1 + 3 z \right].
\end{align}
We integrate over the $k_t^2$ range, equivalent to the $t$ interval $t_0$ to $t_1$ as before, and then demand that the result is equal to the integral over the LO-generated cross section (\ref{eq:LLA}) from $|t_0|$ to $\mu_0$.  We find that the result, equivalent to (\ref{eq:bal}), is
\begin{equation}
z \left[ z^2 + (1-z)^2 \right] \ln\frac{\mu_0^2}{|t_0|} = z \left[( (1 - z)^2 + z^2) \left( \ln\frac{M^2}{|t_0|} + \ln \frac{1 - z}{z}\right)  +  \frac{1 + 3z}{2}(1 - z) -{\cal I}(z,k_0)  \right],
\label{eq:bal2}
\end{equation} 
where again the infrared divergency, as $t_0 \to 0$, cancels; and where we now have a new term 
\begin{equation}
{\cal I}=\Theta (z_\mathrm{max} - z) \left[ 2 (1 + 2z^2 - 2z) \tanh^{-1} \left( \sqrt{1 - \frac{4 z}{(1-z)^2} \frac{k_0^2}{M^2}} \right)  + \frac{1 + 3z}{2} \sqrt{(1 - z)^2 - 4 z\frac{k_0^2}{M^2}} \right] .
\label{eq:I}
\end{equation}
The optimal scale $\mu_0$, determined from (\ref{eq:bal2}), now depends on the choice of the $k_t$-cutoff, $k_0$, due to the presence of ${\cal I}(z,k_0)$.  What is the origin of the theta function in this new term?
Note that the $k_t$-cutoff only has an effect when $k_0$ is smaller than the maximum $k_t$. Thus, there is a maximum value of $z$ for which the cutoff is to be applied, given by
\begin{equation}
z_\mathrm{max} = 1 + 2 \frac{k_0^2}{M^2} - 2 \sqrt{\frac{k_0^2}{M^2}\left(\frac{k_0^2}{M^2}+1\right)}.
\end{equation}

In order to determine the optimal scale $\mu_0(k_0)$, we proceed as before, and integrate (\ref{eq:bal2}) over the parton-parton flux $\dd z/z$. Since the $gq\to q\gamma^*$ subprocess is described by the same diagram as that corresponding to the last step of the LO DGLAP evolution, the scale $\mu_0$ that is calculated in this way provides an exact `resummation' of the most dangerous\footnote{That is terms enhanced by the large value of $\ln(1/x)$.}
%in the incoming parton distributions all the large 
double logarithmic terms of the form $C_n(\alpha_s\ln q^2\ln(1/x))^n$ inside the incoming parton distributions.  After the integration over the flux\footnote{In order to account better for the $x$-dependence of the incoming parton densities we may include in the flux an additional factor $z^\delta$. However, integrating over such a flux factor with $\delta \lapproxeq 0.2$ gives a negligible change in the resulting value of $\mu_0/M$ for the region of interest $k_0 \lapproxeq M/2$.}, 
we find $\mu_0(k_0)$ is now given by
\begin{equation}
\ln\frac{\mu_0}{M} = \frac{3}{4} \left(\frac{1}{2} - I \right),
\end{equation}
in the place of (\ref{eq:nocut}). The additional term $I$ can be evaluated analytically. It is given by
\begin{align}
I\left( w\equiv \frac{k_0^2}{M^2} \right) &= \int^{z_{\rm max}}_0 \dd z  ~{\cal I}(z,k_0) \\
&=  \frac{\left( 66 w^3 + 72 w^2 + 18 w + 12 \right) \ln \left[ (w+1)/w \right] - 66 w^2 - 39 w - 4}{18} .
\end{align}
\begin{figure} [htb]
\begin{center}
\includegraphics[height=10cm]{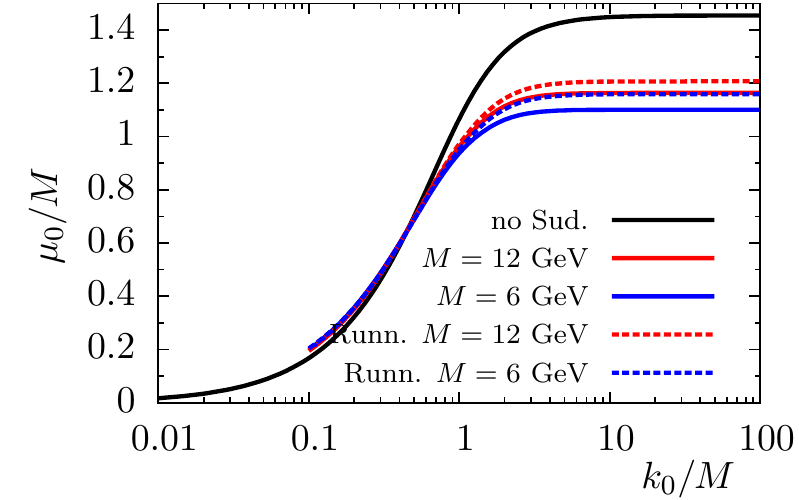}
\vspace{-0.2cm}
\caption{\sf The dependence of the optimal scale $\mu_0$ on the choice to cutoff $k_t<k_0$ on the transverse momentum, $k_t$, of the Drell-Yan $\mu^+\mu^-$ pair of mass $M$. The curves for $k_0/M>0.1$ show how including the Sudakov factor, with and without running $\alpha_s$, changes the dependence of the optimal scale $\mu_0$ on the cutoff $k_0$.}
\label{fig:w}
\end{center}
\end{figure} 
The dependence of the optimal scale, $\mu_0$, on the $k_t$ cutoff, $k_t<k_0$, is shown by the continuous curve in Fig. \ref{fig:w}. In particular, for a cutoff $k_0=M/4$ we find that the optimal scale is $\mu_0=0.39M$, while for $k_0=M/2$ we have $\mu_0=0.70M$.
These values are a considerable improvement on the result $\mu_0=1.45M$ in the absence of the cut, but are somewhat larger than the naive estimate of $\mu_0=k_0$ since, (i) the scale $\mu_0$ is defined in terms of the {\em virtuality} $|t|=k^2_t/(1-z)>k^2_t$, and, (ii), in the $gq \to q\gamma^*$  cross section, (\ref{eq:exact0}), we have a `backward scattering' contribution with
$|t|>|u|$, but where the transverse momentum $k^2_t=tu/\hat s$ is still small.

So it looks as if we have achieved our aim. By selecting data satisfying a cutoff $k_t(\mu^+\mu^-)<k_0$, we do indeed significantly reduce the optimal scale, provided the cutoff satisfies $k_0\lapproxeq M$. But, first, we must check whether allowing for Sudakov effects will alter this conclusion.

\subsection{Including Sudakov effects}

As was shown above, 
it is straightforward to calculate the NLO Drell-Yan cross section accounting for the $k_t<k_0$ cut in the NLO contribution, and to use the scale $\mu_F=\mu_0(k_0)$ in the LO part. However, when
$k_0 \ll M$ we have to take care of the resummation of possible double logarithmic terms, $(\alpha_s\ln^2(M/k_0))^n$, which account
for the small probability not to emit additional `bremsstrahlung'
gluons with $k_t$ larger than $k_0$. In the double Log approximation the
corresponding Sudakov form factor reads
\be
\label{eq:sud} 
T_q=\exp(-\alpha_s S_q(\mu_0,\mu_T))
\ee
with
\be
\label{eq:ST}
S_q=\frac{C_F}{\pi}\ln^2(\mu_T/\mu_0)\ ,
\ee
where $C_F=4/3$. (For an incoming gluon the corresponding expression (\ref{eq:ST}) contains the colour factor $C_A=N_c=3$ instead of $C_F$.) The upper limit $\mu_T$ of the $k_t$ integration in the $T^2$ factor is typically chosen\footnote{It may be more collinear-safe to choose $\mu^2_T=M^2+k_t^2$. However, this is beyond the double log accuracy used in (\ref{eq:sud}) and (\ref{eq:ST}). Recall that the $T$ factor is introduced to resum corrections enhanced by ln$(M/k_t)$ for the case when $k_t \ll M$. Therefore we prefer to simplify the formula by including the remaining corrections in the usual higher ${\cal O}(\alpha_s)$ contributions $C_{\rm rem}^{\rm NLO,NNLO}$, see (\ref{eq:Sud-LO}).} 
to be $\mu_T = M$.  This value provides the correct single logarithmic term in $T$-factor~\cite{FT}. 
Moreover, to be precise,  expression (\ref{eq:ST}) is actually  of the form
\be
\label{eq:ST2}
S_q=\frac{C_F}{4\pi}\ln^2(-\mu^2_T/\mu^2_0) =\ \frac{C_F}{\pi}[\ln(\mu_T/\mu_0)\ +\ i\pi/2]^2\ .
\ee
So we include the $\pi^2$-enhanced term $-C_F\pi/4$ in $S_q$. As was discussed in~\cite{LMRT}, this term arises from soft gluon emission and can be exponentiated\footnote{The imaginary part is cancelled between the amplitude $A$ and its complex conjugate $A^*$.}.

Certainly, the LO contribution should be multiplied by the $T^2$ factor given above. On the other hand, if one takes the complete NLO contribution, the first $\alpha_s$ term of $T$-factor  was already
included in the NLO loop calculation. Therefore, to avoid double counting, we must compensate for this term. That is,
we replace the NLO contribution $C^{\rm NLO}$ by
\be
\label{eq:Sud-LO}
C^{\rm NLO}_{ab}~\rightarrow ~C^{\rm NLO}_{ab} 
~+~\CLO_{ab}~\alpha_s [S_a(\mu_0,\mu_T)+S_b(\mu_0,\mu_T)]\ .
\ee
Note that for the subprocess $qg\to q+\gamma^*$ (that is, for $a=q$ and $b=g$) we allow for the different probabilities of bremsstrahlung from gluon and quark lines. 
We choose the upper scale in the Sudakov $T$-factor to be the mass of the final parton system, that is $\mu_T=M$
in the LO $q\bar q\to \gamma^*$ case and $\mu^2_T=\hat s$ for the
NLO $qg\to q+\gamma^*$ subprocess.

\begin{figure} 
\begin{center}
\includegraphics[height=8cm]{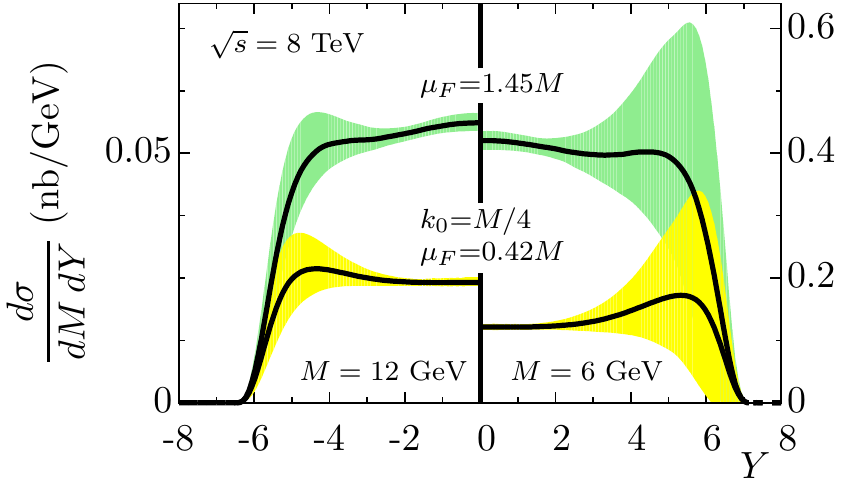}
\vspace{-0.2cm}
\caption{\sf The `LO+NLO' cross sections for the Drell-Yan production of a $\mu^+\mu^-$ pair of mass $M=6$ GeV and $M=12$ GeV, as a function of its rapidity at the LHC energy of 8 TeV. The lower set of curves correspond to the imposition of a cut $k_t(\mu^+\mu^-)<M/4$, where the optimal scale is given by the curves in Fig. \ref{fig:w} with the Sudakov effects included, namely $\mu_0=0.42M$. The upper set of curves correspond to the absence of a $k_t$ cutoff, in which case the optimal scale $\mu_0=1.45M$, see Section \ref{sec:2}. The 1$\sigma$ error corridors obtained using the MSTW NLO parton set are shown.   The scales in the left and right hand sides of the figure have been adjusted to allow  for the naive $1/M^3$ behaviour of the cross section, so that any departure from the `symmetry' of the plot is due to the PDFs.}
\label{fig:uncert}
\end{center}
\end{figure} 
Formally working to NLO accuracy we have to keep the QCD coupling $\alpha_s$ fixed and to use the double Log approximation (\ref{eq:ST}). On the other hand, at relatively low scales the effect of the $\alpha_s$ running may be not negligible. The effect is shown by the dashed curves in Fig. \ref{fig:w}, which are obtained by  using a more precise expression (\ref{eq:tq}), and the analogous expression for incoming gluons. This leads to a small increase in the values of $\mu_0$ for $k_0\gapproxeq M$.  Moreover, it is clear from Fig. \ref{fig:w} that in the region of interest, $k_0$ in the interval $(M/8,~M)$, the possibility of additional QCD radiation (the Sudakov effect) does {\it not} change the value of $\mu_0$ too much.  Recall also that the variation of the upper scale $\mu_T$
corresponds to emission of a hard gluons. We cannot justify the exponentiation of this contribution. It should be considered as a  higher $\alpha_s$ order correction and should be accounted for when calculating the NNLO terms.

\begin{figure} 
\begin{center}
\includegraphics[height=8cm]{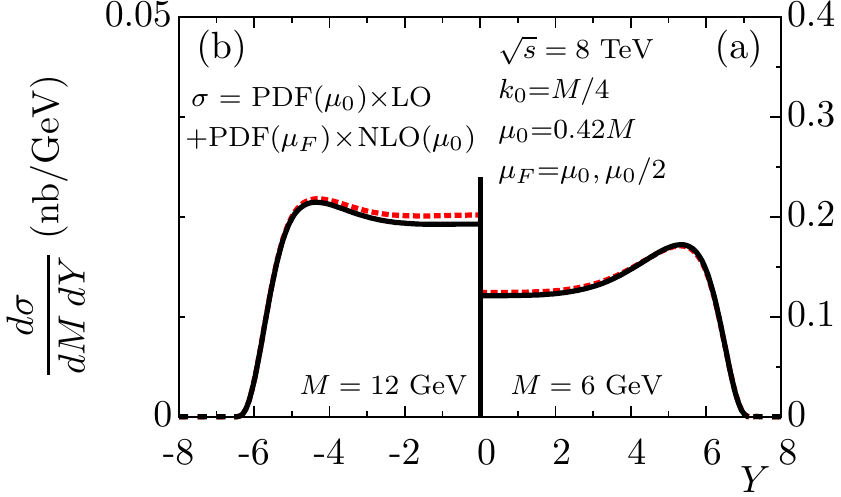}
\vspace{-0.2cm}
\caption{\sf The sensitivity of the Drell-Yan $\mu^+\mu^-$ cross section with respect to variations $\mu_F=\mu_0/2,~\mu_0$ in the scale of the PDFs convoluted with $\CNLO_{\rm rem}$. The cutoff $k_t(\mu^+\mu^-)<M/4$ is imposed, with optimal scale $\mu_0=0.42M$ as given by Fig. \ref{fig:w}.}
\label{fig:muFdep}
\end{center}
\end{figure}

\section{Predictions for low-mass Drell-Yan production  \label{sec:4}}
The cross sections expected using MSTW08 NLO partons are 
plotted as the function of $Y$ in Fig. \ref{fig:uncert} for $M=6$ and 12 GeV for the LHC energy $\sqrt{s}$=8 TeV.
The upper curves correspond to the cross section obtained without the imposition of the $k_t$ cut (and correspondingly without the $T$-factor, i.e. $T=1$), while the lower curves are calculated imposing a cutoff $k_t<M/4$ on the transverse momentum of the produced $\mu^+\mu^-$ system.
For the LO part the {\it optimal} scales $\mu_0=1.45M$ (without the cut) and $\mu_0=0.42M$ (with the cut) were used, as deduced in Sections \ref{sec:2} and \ref{sec:3} respectively. 
%The dashed lines show the effect of variation of the factorization scale $\mu_F=\mu_0/2$ in the remaining NLO contribution.
%As seen from Fig. \ref{fig:uncert}, the use of the optimal scale $\mu_0$ in the LO part provides a sufficient stability with respect to the factorization scale variations in the remaining NLO contribution.

\begin{figure} 
\begin{center}
\includegraphics[height=8cm]{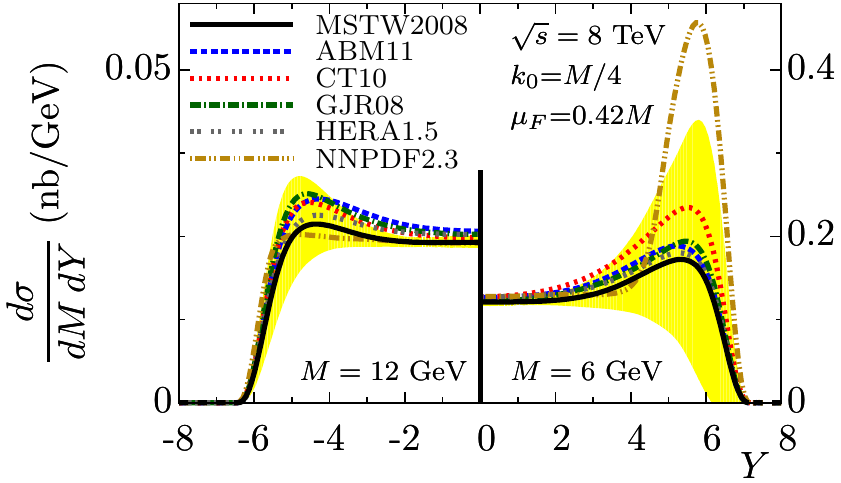}
\vspace{-0.2cm}
\caption{\sf The sensitivity of the Drell-Yan $\mu^+\mu^-$ cross section with respect to the choice of PDFs \cite{MSTW,others}. The cutoff $k_t(\mu^+\mu^-)<M/4$ is imposed, with optimal scale $\mu_0=0.42M$ as given by Fig. \ref{fig:w}.}
\label{fig:pdfs}
\end{center}
\end{figure} 
\newpage
Figs. \ref{fig:muFdep} and \ref{fig:pdfs} show, respectively, the sensitivity of the Drell-Yan  cross section, for the production of a $\mu^+\mu^-$ pair of mass $M$=6 or 12 GeV and rapidity $Y$, on the factorization scale $\mu_F$ and on the choice of NLO PDFs \cite{MSTW,others}, together with the error corridor obtained using the MSTW partons.

In Fig. \ref{fig:muFdep} we plot the cross sections with the cutoff $k_t<M/4$ calculated using the scale $\mu_F=\mu_0$ in the LO part and the factorization scale $\mu_F=\mu_0$ (solid lines) or $\mu_F=\mu_0/2$ (dashed lines) in the remaining NLO part. Just as for the cross sections without the cutoff, shown in Fig. \ref{fig:muF}, 
the use of the optimal scale $\mu_0$ in the LO part provides sufficient stability with respect to the variation of the factorization scale in the remaining NLO contribution.
 We do not consider here the domain of  large $\mu_F>\mu_0>k_0$ in the remaining NLO contribution, since then a large transverse momentum  of the incoming parton immediately violates the cutoff condition $k_t<k_0$.

\section{Conclusion  \label{sec:5}}

It is well known that the Drell-Yan cross section for  {\it low-mass} $\mu^+\mu^-$ production depends sensitively on the choice of factorization scale, $\mu_F$.  In an earlier publication \cite{OMR} we showed how this sensitivity can be greatly reduced by fixing the scale $\mu_F$ 
by comparing the exact 
contribution of  NLO $gq\to q\gamma^*$ subprocess  with the analogous 
contribution generated by LO DGLAP evolution.
Since the NLO $gq\to q\gamma^*$ subprocess is described by the same Feynman diagram as that in the last step of the evolution, the scale $\mu_0$ at which the LO evolution reproduces the exact NLO result may be treated as the factorization scale probed by the Drell-Yan process.
After
the scale $\mu_0$ is fixed in the LO part, the prediction is found to depend weakly on 
the variation of factorization scale in the remaining NLO part.
In Section \ref{sec:2} we illustrated the procedure by calculating the value of the optimal scale analytically, and found $\mu_0=1.45M$, where $M$ is the mass of the $\mu^+\mu^-$ pair. Experiments have so far reached down to $M \sim 6$ GeV and up to rapidities $Y \sim 4$, which probe parton PDFs as low as $x \sim 10^{-5}$, albeit at high $Q^2$ scales, $Q^2=(1.45M)^2 \sim 75~\GeV^2$ for $M\sim 6$ GeV.

In this paper, we show that lower scales can be probed if we impose a cutoff on the transverse momentum of the $\mu^+\mu^-$ pair, $k_t<k_0$. In this case the optimal scale depends on the choice of the cutoff $k_0$. First, in Section \ref{subsec:3.1}, we performed the analytic calculation of $\mu_0(k_0)$, see Fig. \ref{fig:w}. Then, we included the effects of the Sudakov suppression $T$-factor. Fig. \ref{fig:w} shows that in the region of not too low scales (corresponding to $k_0\gapproxeq M/8$) the $T$-factor makes only a small increase in the value of the optimal scale. For example if we take $k_0=M/4$, then the Sudakov effect only increases $\mu_0$ from $0.39M$ to $0.42M$. Hence $M \sim 6$ GeV data probe PDFs at $Q^2\sim 6~\GeV^2$.  The cost of using this cutoff in shown in Fig. \ref{fig:uncert}: that is, imposing the cut $k_t<M/4$ GeV on the $M=6$ GeV data,  means we keep only about a third of the events (and for $M=12$ GeV we keep about half of the events, but really, to reach lower scales, we should now choose a lower cutoff, say $k_0 \sim M/8$)

Fig. \ref{fig:uncert} is computed using MSTW NLO partons \cite{MSTW}, and shows the error corridors of the predictions. As expected the percentage uncertainty is largest at the lower $Q^2$ scales, which indicates the value of imposing the cutoff to reveal information of the low $x$ behaviour of PDFs.
Fig. \ref{fig:pdfs} shows the range of predictions obtained using other recent sets of PDFs \cite{others}.

What is the impact of such measurements on global PDF analyses? Clearly, the measurements can probe a small $x$ domain so far unexplored by data.  If, at such small $x$, we were to believe in pure DGLAP evolution, then the majority of the low $x$ quarks (antiquarks) comes
from the $g\to q\bar q$ splitting. Therefore the $M$ dependence of the Drell-Yan cross section will probe the gluon distribution in a similar way to how the scaling violations in deep inelastic scattering data,  
$dF_2(x,Q^2)/dQ^2$, probe, via DGLAP evolution, the low $x$ gluon density.
Note that for usual fully inclusive kinematics, at a fixed initial energy, changing the value of $M$  simultaneously changes the value of $x$. Now, by selecting events with different $k_t$ cuts, we may vary the scale $\mu_F=\mu_0$ while keeping the value of $x$ fixed.

Figs. \ref{fig:uncert} and \ref{fig:pdfs} may give a misleading impression about the value of low-mass Drell-Yan data.  Data for the production of $\mu^+\mu^-$ pairs of mass $M \lapproxeq 20$ GeV at rapidities $Y \gapproxeq 3$ probed the unexplored domain $x \lapproxeq 10^{-4}$. In this $x$ region the PDFs curves in Fig. \ref{fig:pdfs} are based on unreliable extrapolations assuming pure DGLAP forms.  Clearly the extrapolations must be modified to allow for  absorptive effects and  ln$(1/x)$ corrections, etc. Indeed, it is fair to say that there is no meaningful prediction in this domain. Rather we should regard the low-mass Drell-Yan data at the LHC as offering an important {\it direct measurement} of the PDFs in this unexplored low $x$ region.

\section*{Acknowledgements}
EGdO and MGR thank the IPPP at the University of Durham for hospitality. 
 This work was supported by the grant RFBR 11-02-00120-a
and by the Federal Program of the Russian State RSGSS-4801.2012.2;
and by FAPESP (Brazil) under contract 2012/05469-4.

\end{document}